\documentclass[a4paper,11pt]{article}
\usepackage{pos}
\usepackage{graphicx,amsmath,amsfonts}
\usepackage{wrapfig}
\newcommand{\inspire}[1]{[\href{https://inspirehep.net/literature?q=#1}{\sc inSPIRE}]}
\usepackage{supertabular}
\usepackage{slashed}
\usepackage{multirow,multicol,diagbox,array} 

\allowdisplaybreaks[2]
\usepackage{calrsfs}
\DeclareMathAlphabet{\pazocal}{OMS}{zplm}{m}{n}

\def\ds{\displaystyle}
\usepackage{graphicx,amsmath,amsfonts}
\usepackage{wrapfig}
\usepackage{caption}

\title{Internal structure of exotic hadron candidate $f_0$(980) \\
by using fragmentation functions}

\author[a,b,c]{S. Kumano}
\affiliation[a]{Quark Matter Research Center,
    Institute of Modern Physics, Chinese Academy of Sciences,\\
    Lanzhou, 730000, China}
\affiliation[b]{ Southern Center for Nuclear Science Theory,
    Institute of Modern Physics, Chinese Academy of Sciences,\\
    Huizhou, 516000, China}
\affiliation[c]{KEK Theory Center, Institute of Particle and Nuclear Studies, KEK,
    Oho 1-1, Tsukuba, 305-0801, Japan}
\emailAdd{kumanos@impcas.ac.cn}

\abstract{
High-energy hadron reactions could be appropriate to find
exotic evidences in exotic hadron candidates instead of
global observables such as masses, spins, parities, and decay widths,
because quarks and gluons are explicit degrees of freedom.
One of possible methods is to use fragmentation functions (FFs)
by taking advantage of properties on favored and disfavored functions,
which corresponds to valence-quark and sea-quark distributions
in parton distribution functions. Looking at these FFs, we should
be able to find the exotic nature as the valence-quark distributions
reflect the nature of valence constituents as shown in the pion and the proton.
Recent accurate measurements of the $f_0$(980) FFs by
the Belle collaboration made it possible to find its internal
configuration by looking at their second moments and functional forms.
Global analysis results of the $f_0$(980) FFs indicate
that its internal configuration looks like $s\bar s$,
which is different from a tetraquark or $K\bar K$-molecule like configuration
suggested from low-energy studies. 
This fact indicates that the internal configuration looks different
depending on the energy. At low energies, it looks like a tetraquark 
($K\bar K$ molecule) hadron, but it looks like a $q\bar q$ hadron
at high energies. In the similar way, some exotic hadron candidates 
could become ordinary $q\bar q$ or $qqq$ hadrons at high energies,
although they are interpreted as exotic at low energies.
This kind of new idea should be tested by future experiments
especially by looking at energy or momentum dependencies
in high energy hadron reactions.
}

\FullConference{33rd International Workshop on Deep Inelastic Scattering 
and Related Subjects (DIS2026)\\
4-8 May 2026\\
Bologna, Italy\\}

\begin{document}
\maketitle

\ \\[-1.75cm]
\section{Introduction}
\label{introduction}
\vspace{-0.15cm}

The hadrons which do not fit into the category of the naive-quark-model
configurations $q\bar q$ and $qqq$ are called exotic hadrons.
Now, there are many indications of exotic hadron candidates
especially in heavy-quark systems. On the other hand, there are 
controversial hadrons for a long time in the mass region of 1 GeV,
and typical examples are $f_0 (980)$, $a_0 (980)$, and $\Lambda (1405)$
\cite{pdg-2024}.
In general, it is not easy to show undoubtful exotic evidences 
from global observables like masses, spins, parities, and decay widths
\cite{cik-1993, sk-2015}.
Alternatively, high-energy reactions could be appropriate
to indicate the exotic nature because explicit degrees of freedom, 
quarks and gluons, are basic constituents of quantum chromodynamics (QCD).
Various possibilities were proposed for finding
exotic signatures at high energies
by using parton distribution functions, 
generalized parton distributions \cite{gpd}, constituent counting rule 
\cite{const-conting-2013,const-conting-2016}, 
and fragmentation functions (FFs) \cite{hkos-2008,sk-ffs-2025}.

In this work, the FFs are used for finding the internal structure
of $f_0(980)$ \cite{sk-ffs-2025}. 
According to the basic quark model, the light scalar meson $f_0$(980) with
$J^{PC}=0^{++}$ is the $^3 P_0$ quarkonium with the flavor
structure $f_0 =(u\bar u+d\bar d)/\sqrt{2}$ \cite{pdg-2024}. 
However, the strong decay width $\Gamma(f_0 \rightarrow \pi\pi)$ is
too large if such a $q\bar q$ configuration is used for $f_0$(980)
to explain the experimental width \cite{f0width-th}.
On the other hand, a sizable strange-quark component in $f_0 (980)$
was suggested by the Fermilab-E791 collaboration 
in the decay $D_s^+ \rightarrow \pi^- \pi^+ \pi^+$ through 
$D_s^+ \rightarrow f_0 (s\bar s) \pi^+$ with $s\bar s$ quarks in $f_0 (980)$.
In addition, because the $f_0$ mass is below the $K\bar K$ threshold, 
there is a possibility of a $K\bar K$ molecule or 
a tetraquark state $f_0=(u\bar u s\bar s+d\bar d s\bar s)/\sqrt{2}$
as suggested in the MIT bag model. 
QCD-sum-rule and lattice QCD studies support the tetraquark
($K\bar K$) configuration; however, the glueball possiblity 
$f_0 =gg$ was ruled out in the  1 GeV region.

As proposed in Ref.\,\cite{hkos-2008},
the FFs could be used for probing the exotic nature of hadrons.
As the PDFs are classified by the valence-quarks and sea-quarks,
the FFs have favored and disfavored functions.
The favored fragmentation indicates that the initial paton
exists in a hadron as a main constituent, whereas the disfavored one
is the fragmentation from a sea constituent.
The FF method is appropriate for finding the internal structure
of exotic hadrons because they could be produced in timelike processes
$e^+ +e^- \to h +X$. Because the exotic hadron candidates are 
unstable particles, they cannot be used as fixed targets,
and hence it is not easy to investigate them directly in spacelike processes,
although semi-inclusive production processes can be investigated.
If a parton is a main constituent in a hadron, its FF
should be larger than other disfavored FFs. Namely,
the second moments of the favored FFs should be much
larger than the second moments of the disfavored FFs. 
In addition, the favored FFs should be peaked in the relatively large-$z$
region because more energy is transferred to $f_0 (980)$
from the initial parton, where as the the disfavored FFs should be
distributed mainly in the small-$z$ region.

In the previous analysis \cite{hkos-2008}, the $f_0 (980)$ FF measurements 
were not accurate enough to probe its internal structure.
However, the Belle measurements of 2025 are accurate 
to reinvestigate this topic \cite{KEKB-2025}. 
The Belle data were taken at the center-of-mass (c.m.) energy 10.58 GeV 
and other data were mainly at the $Z$-mass 91.2 GeV. 
The large gap between these energies makes it possible 
to find the scaling violation of the FFs, namely
for fixing the gluon FF of $f_0 (980)$ with reasonable accuracy.
Then, the determination of quark FFS should become easier 
than the previous analysis for finding the internal structure.
In this paper, the results of the recent analysis \cite{sk-ffs-2025}
are discussed.
First, the analysis method is explained
for calculating the fragmentations of $f_0 (980)$
in Sec.\,\ref{formalism}. Then, numerical results are shown 
in Sec.\,\ref{results} and
they are summarized in Sec.\,\ref{summary}.


\section{Analysis method}
\label{formalism}

In the reaction $e^+ +e^- \to h +X$, the FFs are expressed by 
the variable $Q^2$ ($=q^2$) with the virtual-photon or $W$ momentum $q$
and  the variable $z$ defined by the hadron energy $E_h$ 
and the center-of-mass energy $\sqrt{s}$ ($=\sqrt{Q^2}$) as
$z \equiv E_h/(\sqrt{s}/2)$.
They are defined by the hadron-production
cross section and the total hadronic cross section $\sigma_{tot}$ as
\cite{esw-book,FFs-global-1,FFs-global-2} 
\begin{equation}  
F^h(z,Q^2) = \frac{1}{\sigma_{tot}} 
\frac{d\sigma (e^+e^- \rightarrow hX)}{dz} .
\label{eqn:def-ff}
\end{equation}
Experimental data are shown by the form of Eq.\,(\ref{eqn:def-ff})
or slightly modified ones.
Theoretically, the fragmentation function $F^h(z,Q^2)$ is given 
by the sum of primary-parton fragmentation contributions as
\begin{equation}  
F^h(z,Q^2) = \sum_i C_i(z,\alpha_s) \otimes D_i^h (z,Q^2),
\label{eqn:def-ffqqbarg}
\end{equation}
where $\otimes$ indicates the convolution integral
$f (z) \otimes g (z) = \int^{1}_{z} dy f(y) g(z/y)/y$,
$C_i(z,\alpha_s)$ is a coefficient function with
the running coupling constant $\alpha_s$, and
$D_i^h(z,Q^2)$ is the fragmentation function
of the hadron $h$ from the parton $i$ ($=u,\ d,\ s,\ \cdot\cdot\cdot,\ g$).

The total cross section is given by the contributions from
the $q\bar q$-pair creation processes, 
$e^+e^- \rightarrow \gamma \rightarrow q\bar q$ and
$e^+e^- \rightarrow Z \rightarrow q\bar q$, with higher-order
corrections as
\begin{equation}   
\sigma_{tot}=\sum_q \sigma_0^{q} (s)
   \left [ 1 + \frac{\alpha_s(Q^2)}{\pi} \right ] .
\end{equation}
Here, the perturbative correction is included up to the 
next-to-leading order (NLO) of $\alpha_s$.
The cross section for producing a $q\bar q$ pair is given by
\begin{align}
\sigma_0^{q} (s) = \frac{4 \pi \alpha^2}{s} \, 
  [ \, e_q^2+2 e_q c_V^e c_V^q \rho_1(s) 
   + (c_V^{e\, 2}+c_A^{e\, 2})(c_V^{q\, 2}+c_A^{q\, 2}) \rho_2(s) \, ].
\end{align}
The $\rho_1(s)$ and $\rho_2(s)$ terms come from $\gamma$-$Z$
interference and $Z$ processes and they are
\begin{align}
\rho_1(s) = & \frac{1}{4\sin^2\theta_W\cos^2\theta_W} 
       \frac{s(M_Z^2-s)}{(M_Z^2-s)^2+M_Z^2\Gamma_Z^2}, 
\nonumber \\
\rho_2(s) = & \left(\frac{1}{4\sin^2\theta_W\cos^2\theta_W}\right)^2
       \frac{s^2}{(M_Z^2-s)^2+M_Z^2\Gamma_Z^2},
\end{align}
where $\alpha$ is the fine structure constant in quantum electrodynamics
(QED), $e_q$ is a quark charge, 
$M_Z$ and $\Gamma_Z$ are the mass and width of $Z$, respectively,
and $\theta_W$ is the weak-mixing angle.
The vector and axial-vector couplings $c_V^f$ and $c_A^f$ of a fermion $f$
are given by the third component of the weak isospin $T_f^3$ and
the fermion charge $e_f$ as
\begin{equation}
c_V^f = T_f^3 -2 e_f \sin^2 \theta_W, \ \ \ 
c_A^f = T_f^3 .
\end{equation}
Therefore, their expressions in terms of $\sin^2 \theta_W$ are
$c_V^e=-\frac{1}{2}+2\sin^2 \theta_W$ and $c_A^e=-\frac{1}{2}$
for the electron, 
$c_V^u=+\frac{1}{2}-\frac{4}{3}\sin^2 \theta_W$ and $c_A^u=+\frac{1}{2}$
for up, charm, and top quarks, and
$c_V^d=-\frac{1}{2}+\frac{2}{3}\sin^2 \theta_W$ and $c_A^d=-\frac{1}{2}$
for down, strange, and bottom quarks.

Next, the FFs should be prepared with appropriate parameters
for the global analysis 
\cite{esw-book,FFs-global-1,FFs-global-2} of $f_0$(980) data
\cite{hkos-2008,sk-ffs-2025}.
In Sec.\,\ref{introduction}, the following possible configurations
\cite{cik-1993}
\begin{equation}
f_0 (980) = \frac{u\bar u+d\bar d}{\sqrt{2}}, \ \ 
s\bar s, \ \ 
\frac{u\bar u s\bar s+d\bar d s\bar s}{\sqrt{2}} \ (K\bar K), \ \ 
gg ,
\end{equation}
were discussed for $f_0$(980), although the last glueball
possibility was ruled out by the lattice QCD.
Considering all of these possibilities, we find that 
light-quark FFs should be identical 
at the initial scale $Q_0^2$,
other heavy-quark and -antiquark functions should be the same,
and the gluon function 
needs to be separately given.
These functions are expressed by parameters, 
which are determined by a $\chi^2$ analysis of the data for
$e^++e^- \rightarrow f_0+X$. 
In the current analysis, each function is expressed 
by the three parameters
$N$, $\alpha$, and $\beta$ as
\begin{align}
D_{u}^{f_0} (z,Q_0^2) & = D_{\bar u}^{f_0} (z,Q_0^2)
 = D_{d}^{f_0} (z,Q_0^2) =  D_{\bar d}^{f_0} (z,Q_0^2)
 = N_{u} z^{\alpha_u} (1-z)^{\beta_u} ,
\nonumber \\
D_{s}^{f_0} (z,Q_0^2) & = D_{\bar s}^{f_0} (z,Q_0^2) 
 = N_{s} z^{\alpha_s} (1-z)^{\beta_s}, 
\nonumber \\
D_{c}^{f_0} (z,m_c^2) & = D_{\bar c}^{f_0} (z,m_c^2) 
 = N_{c} z^{\alpha_c} (1-z)^{\beta_c},
\nonumber \\
D_{b}^{f_0} (z,m_b^2) & = D_{\bar b}^{f_0} (z,m_b^2) 
 = N_{b} z^{\alpha_b} (1-z)^{\beta_b},
\nonumber \\ 
D_{g}^{f_0} (z,Q_0^2) & 
 = N_{g} z^{\alpha_g} (1-z)^{\beta_g}
\end{align}
The charm and bottom functions are defined at their
mass threshold values in the analysis of $Q_0^2 =1$ GeV$^2$.
$Q^2$ evolutions of the FFs to the experimental points
are calculated in the standard 
DGLAP (Dokshitzer-Gribov-Lipatov-Altarelli-Parisi) equation
in the NLO of $\alpha_s$ \cite{FFs-evol-2012}.
The second moments $M_i^h$ of the functions $D_i^h (z,Q^2)$
satisfy the relation
\begin{equation}
\sum_h M_i^h = \sum_h \int_0^1 dz \, z \, D_i^h (z,Q^2) = 1 ,
\label{eqn:sum}
\end{equation}
where the sum of $h$ should be taken over all the hadrons.
The moment $M_i^h$ is the energy fraction for the hadron $h$ 
which is created from the parton $i$, and 
it is related to the parameter $N_i^h$ by
$N_i^h = M_i^h / B(\alpha_i^h+2, \beta_i^h+1)$,
where $B(\alpha_i^h+2, \beta_i^h+1)$ is the beta function.

\begin{figure}[b]
\vspace{-0.30cm}
\centering
\begin{minipage}[c]{0.50\textwidth}
\footnotesize
\begin{tabular}
{l@{\extracolsep{10ptplus1fil}}c@{\extracolsep{10ptplus1fil}}c
@{\extracolsep{10ptplus1fil}}c@{\extracolsep{10ptplus1fil}}c
@{\extracolsep{10ptplus1fil}}c@{\extracolsep{10ptplus1fil}}c}
Exp.  &  Lab. \ & Ref.  & \hspace{-0.20cm} $\sqrt{s}$\ \, & \hspace{-0.30cm} Data \# \hspace{-0.20cm}
            & Var. & \hspace{-0.20cm} Cross \\[+0.10cm]
\hline
\ \\[-0.20cm]
HRS         &  SLAC   & \cite{hrs86}    & \hspace{-0.20cm} 29 \ \ \ \  & \, \hspace{-0.20cm} 4 
            & $z_p$      
            & \ \ \ \hspace{-0.20cm} $\ds\frac{s}{\beta}\frac{d\sigma}{dz}$  \\[+0.40cm]
DELPHI \hspace{-0.25cm}    
            &  CERN   & \cite{delphi9599} & \hspace{-0.20cm} 91.2  \,  &  \hspace{-0.20cm} 11  
            & $z_p$      
            & \hspace{-0.20cm} $\ds\frac{1}{\sigma_{\rm tot}}\frac{d\sigma}{dz_p}$ \\[+0.40cm]
OPAL        &  CERN   & \cite{opal98}   & \hspace{-0.20cm} 91.2  \,    & \, \hspace{-0.20cm} 8 
            & $\bar z_p$ 
            & \hspace{-0.20cm} $\ds\frac{1}{\sigma_{\rm tot}}\frac{d\sigma}{d\bar z_p}$  \\[+0.40cm]
Belle       &  KEK \  & \cite{KEKB-2025}  & \, \hspace{-0.20cm} 10.58 \hspace{0.10cm} & \hspace{-0.20cm} 40  
            & $z_p$ 
            & \ \ \ \ \ \ \hspace{-0.20cm} $\ds\frac{d\sigma}{dz_p}$ \\[+0.40cm]
\hline
total       &         &                 &             &  63 
            &  
            & 
\end{tabular}
\vspace{-0.20cm}
\captionof{table}{Experimental data for $e^+ +e^- \rightarrow f_0 (980) +X$.}
\label{tab:exp-f0}
\end{minipage}
\vspace{+0.20cm}
\hspace{+0.30cm}
\normalsize
\vspace{-0.90cm}
\begin{minipage}[c]{0.46\textwidth}
\begin{center}
   \includegraphics[width=7.0cm]{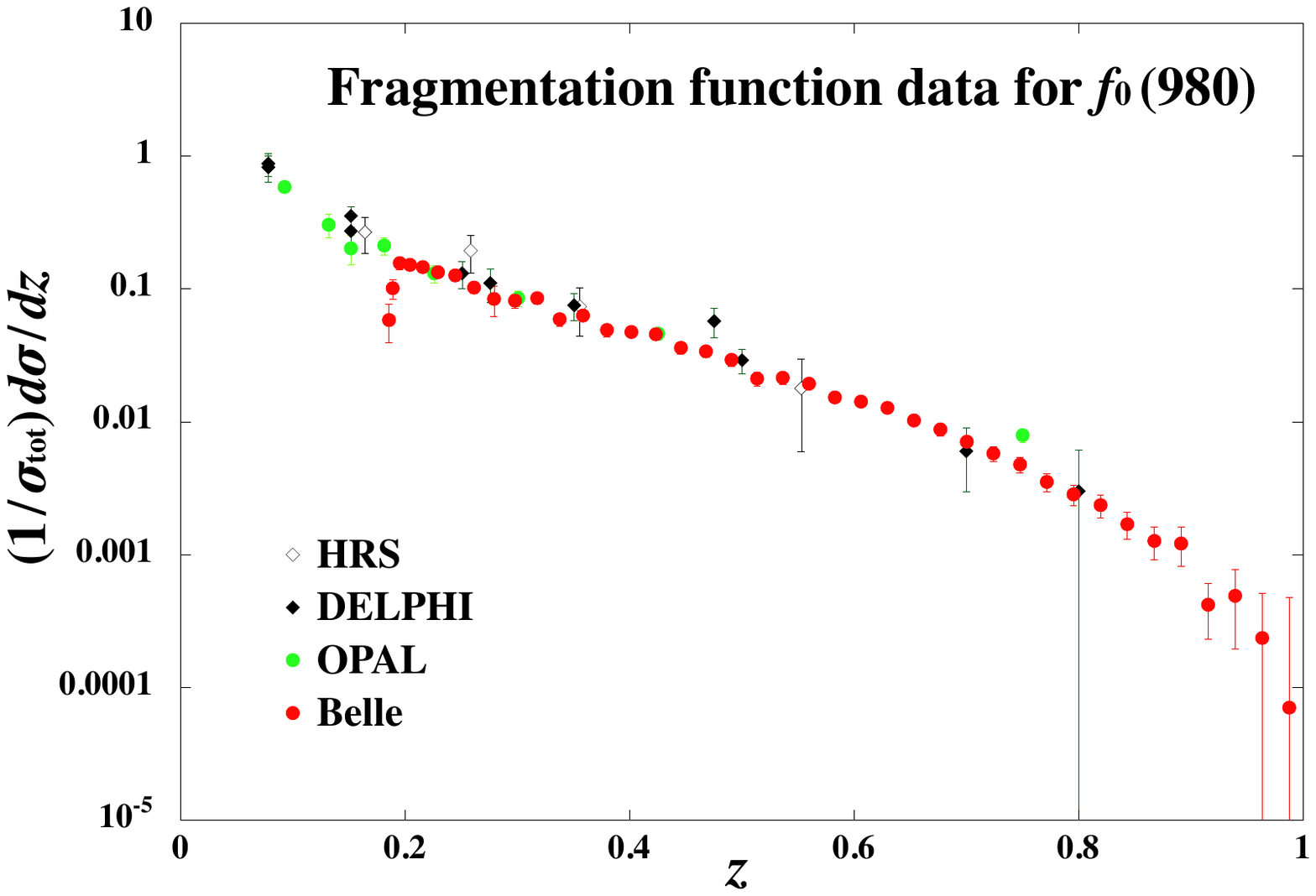}
\end{center}
\vspace{-0.7cm}
\caption{Fragmentation-function data are shown 
for $f_0$(980) by
$F^{f_0} = d\sigma (e^+ e^- \to f_0 X) / dz /\sigma_{\rm{tot}}$.}
\label{fig:f0_980-data}
\end{minipage}
\vspace{+0.20cm}
\end{figure}

Used $e^++e^- \rightarrow f_0+X$ measurements
\cite{KEKB-2025,hrs86,delphi9599,opal98}
are listed in Table \ref{tab:exp-f0}.
Different kinematical variables
are used in showing the data depending on 
the experimental groups:
$ z = z_E = 2 E_h /\sqrt{s}$, 
$ z_p = 2 p_h / \sqrt{s-4 m_h^2}$, 
$ \bar z_p = 2 p_h / \sqrt{s}$,
where $p_h$ and $m_h$ are the $f_0$ momentum and its mass, respectively.
Experimental data are listed by slightly different cross sections, which
are converted to the FFs 
$F^{f_0}(z,Q^2)$ of Eq.\,(\ref{eqn:def-ff})
for the $\chi^2$ analysis, and they are shown 
in Fig.\,\ref{fig:f0_980-data}.
The data number is still not large; however, 
the Belle data are taken at 10.58 GeV c.m. energy,
which is much smaller than 91.2 GeV of the DELPHI and OPAL data.
This energy gap is useful for determining the FFs, especially
on the gluon part.
In Fig.\,\ref{fig:f0_980-data}, experimental data have different 
$Q^2$ values depending on the collaborations; however, all the data
look consistent. Particularly, we notice that the experimental
situation is much improved for $f_0 (980)$ by the the Belle data in 2025.
In the previous global analysis in 2008 \cite{hkos-2008},
only the HRS, DELPHI, and OPAL data were used.
However, they were taken in the limited kinematical points with large errors,
so that it was almost impossible to determine the FFs of $f_0 (980)$ in 2008.
Now, the precise Belle data, taken in the wide kinematical region of $z$,
make it possible to extract the $f_0 (980)$ FFs with reasonable accuracy.

\section{Results}
\label{results}

\begin{figure}[b]
\vspace{-0.20cm}
\centering
\begin{minipage}[c]{0.46\textwidth}
   \includegraphics[width=7.3cm]{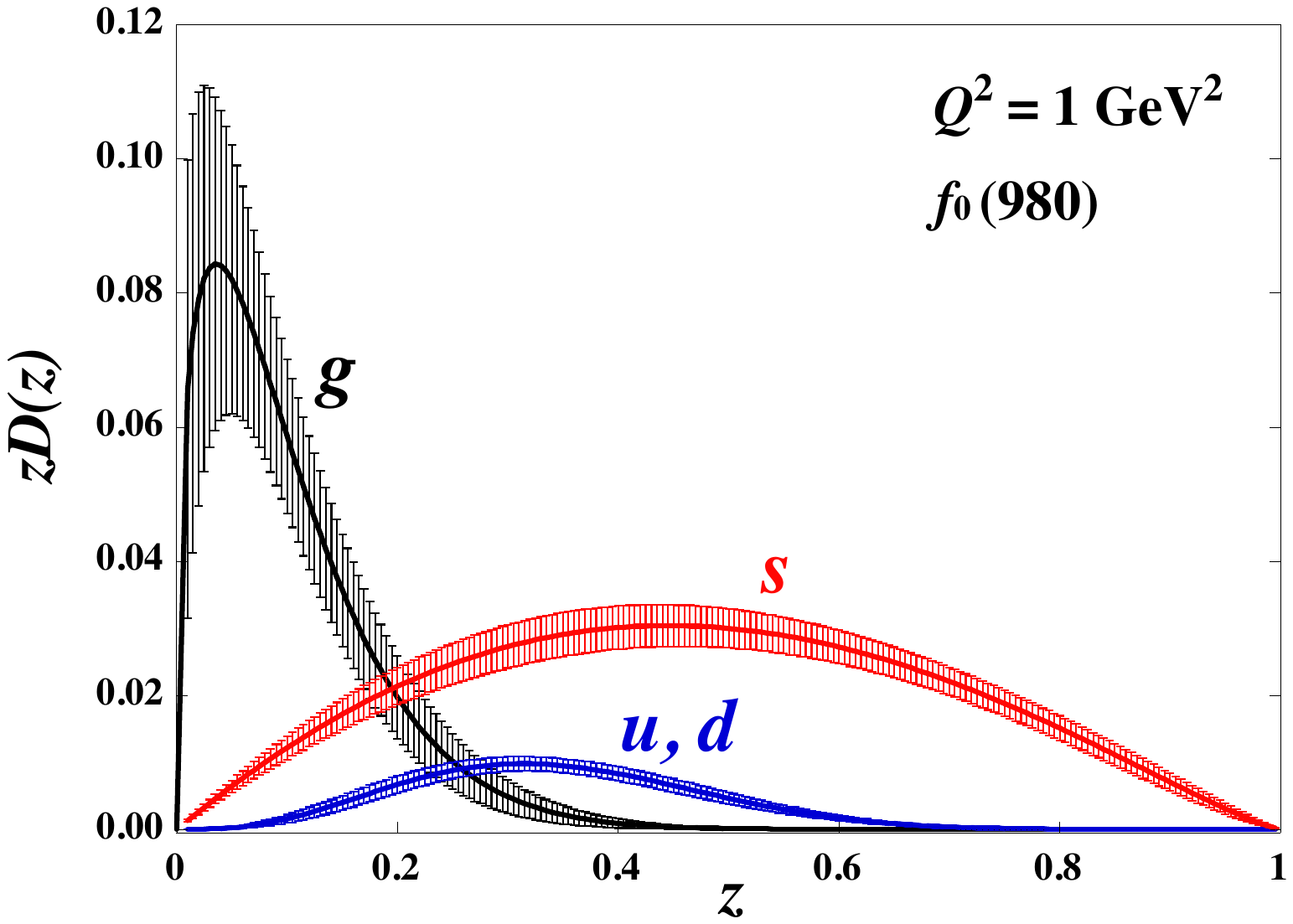}
\vspace{-0.9cm}
\captionof{figure}{Determined fragmentation functions.}
\label{fig:f0_980-ffs}
\vspace{-0.20cm}
\end{minipage}
\hspace{+0.30cm}
\begin{minipage}[c]{0.50\textwidth}
\vspace{-0.22cm}
\footnotesize
\begin{tabular}{ccc}
Configuration   
                        & 2nd moment
                        & Function      \\[+0.10cm]
\hline
\ \\[-0.20cm]
$(u\bar u+d\bar d)/\sqrt{2}$ 
                        & \,\, $M_s<M_u<M_g$ \ \ 
                        & $z_u^{\rm max}>z_s^{\rm max}$   \\[+0.20cm]
$s\bar s$ 
                        & $M_u  <   M_s \lesssim M_g$    
                        & $z_u^{\rm max}<z_s^{\rm max}$   \\[+0.20cm] 
$(u\bar u s\bar s+d\bar d s\bar s)/\sqrt{2}$ 
           & $M_u \sim M_s \lesssim M_g$
           & $z_u^{\rm max} \sim z_s^{\rm max}$   \\[+0.20cm] 
$gg$   
                        & $M_u \sim M_s < M_g$
                        & $z_u^{\rm max} \sim z_s^{\rm max}$   \\[+0.10cm]
\hline
\end{tabular}
\normalsize
\captionof{table}{Possible $f_0(980)$ configurations 
and typical features of its fragmentation functions.}
\label{tab:f0-config}
\vspace{-0.70cm}
\end{minipage}
\end{figure}

In the $\chi^2$ analysis of the data, two parameters $\beta_c$ and $\beta_b$
in the charm and bottom functions are fixed at $\beta_c = \beta_b =10$.
The up-quark (down-quark), strange-quark, and gluon parameters are not fixed
because they are used for judging the internal configuration of $f_0(980)$,
so that the total number of the parameters is 13.
The analysis was done in the NLO, and
the FFs were determined with $\chi^2/\rm{d.o.f.}=1.43$,
and the obtained FFs $D_u^{f_0} (z) =D_d^{f_0}(Z)$, $D_s^{f_0}(z)$, 
and $D_g^{f_0} (z)$ are shown at $Q^2=1$ GeV$^2$ 
in Fig.\,\ref{fig:f0_980-ffs} \cite{sk-ffs-2025}.
Although it was impossible to determine them in 2008, it is now
possible to discuss the FFs of $f_0 (980)$ with the reasonable accuracy.
The function $D_s(z)$ is distributed in the large-$z$ region,
whereas $D_u(z)$ and $D_d(z)$ are in the smaller-$z$ region.
The function $D_s(z)$ is larger than $D_u(z)$ and $D_d(z)$ in magnitude.
The gluon function $D_g(z)$ is distributed in the small-$z$ region.
These second moments are calculated at $Q^2=1$ GeV$^2$,
and they have the relation:
\begin{align}
M_u = M_d = 0.00329 \pm 0.00012 \ll 
M_s = 0.01936 \pm 0.00041 \sim  
M_g = 0.01231  \pm 0.00074 .
\label{eqn:moment-relation}
\end{align}
The other parameters values are
$\alpha_g = -0.559 \pm 0.071$,
$\beta_g  = 11.86  \pm 0.94$,
$\alpha_u = 2.33 \pm 0.11$,
$\beta_u  = 7.266 \pm 0.060$,
$\alpha_s = 0.2623 \pm 0.0023$,
$\beta_s  = 1.276 \pm 0.043$,
$M_c      = 0.00073 \pm 0.00012$,
$\alpha_c = 3.557 \pm 0.073$,
$M_b      = 0.000983 \pm 0.000074$, and
$\alpha_b = 14.53 \pm 0.074$.

From these analysis results, we discuss the possible internal
configuration of $f_0$(980). 
The criteria are summarized in Table \ref{tab:f0-config}
by using the second moments of the FFs and their functional forms
\cite{hkos-2008}. 
First, let us explain the $s\bar s$ configuration for $f_0 (980)$
in Table\,\ref{tab:f0-config} as an example.
In this case, a gluon is emitted from a strange quark (or antiquark),
and then the gluon splits into an $s\bar s$ pair
to form the $s\bar s$ bound state $f_0$ together
with the initial strange quark (or strange antiquark).
This is a favored fragmentation process.
On the other hand, the $u$ quark fragmentation to $f_0 (s\bar s)$
is an unfavored fragmentation process because 
the $s \bar s$ from the gluon splitting could form $f_0 (s\bar s)$
by radiating another gluon to have the color singlet state.
In the same way, the gluon fragmentation process to $f_0 (s\bar s)$
is a disfavored one by another gluon radiation.
In the case of $f_0 (s\bar s)$,
the second moment of $D_s(z)$ should be larger because of
the favored fragmentation without the gluon emission.
Furthermore, more energy is transferred to $f_0 (980)$
from the initial $s$ (or $\bar s$), so that
$D_s(z)$ is distributed at large $z$.
The same discussions can be made by assuming the $f_0 (980)$
as $(u\bar u+d\bar d)/\sqrt{2}$, $(u\bar u s\bar s+d\bar d s\bar s)/\sqrt{2}$,
or $gg$. 
From these considerations, specific features in the second moments 
and the $z$-dependent functional forms are listed
in Table\,\ref{tab:f0-config}.

The determined moment relation in Eq.\,(\ref{eqn:moment-relation})
suggests that $u$ and $d$ quarks are not
the main components of $f_0 (980)$, and this fact rules out
the configurations $(u\bar u+d\bar d)/\sqrt{2}$ and 
$(u\bar u s\bar s+d\bar d s\bar s)/\sqrt{2}$ (or $K\bar K$).
In fact, the FFs in Fig.\,\ref{fig:f0_980-ffs} indicate
that $D_s(z)$ is much larger than $D_{u,d}(z)$ in magnitude.
In contrast, the $u$ and $\bar d$ 
are the main components in $\pi^+$ 
in the recent global analysis 
(J. Gao {\it et al}., 2025 \cite{FFs-global-2}),
as shown by the second moments,
$M_{u,\bar d}^{\pi^+} = 0.481 \gg  M_g^{\pi^+} = 0.102,
\ M_{s,\bar s}^{\pi^+} = 0.234, M_{d, \bar u}^{\pi^+} = 0.119$
in the NLO analysis at $Q^2=(1.4)^2$ GeV$^2$.
The gluon function $D_g(z)$ itself is as large as $D_s(z)$;
however, it is distributed in the small-$z$ region, which
suggests that the gluon is not the main component as
a constituent particle.
From the criteria of Table\,\ref{tab:f0-config} together with
Fig.\,\ref{fig:f0_980-ffs} and Eq.\,(\ref{eqn:moment-relation}),
the appropriate internal configuration of $f_0 (980)$
should be 
\vspace{-0.20cm}
\begin{align}
f_0 (980) =  \ s\bar s \ \ \ \text{from the global analysis of its FFs},
\nonumber
\vspace{-0.00cm}
\end{align}
\\[-0.80cm]
because the $s$-quark moment $M_s$ is much larger than
the $u$- and $d$-quark ones $M_{u,d}$
and because the $D_s(z)$ is distributed in 
the larger-$z$ region than the other FFs.
Even though the global analysis is repeated by changing 
the initial scale is changed for $Q_0^2=4$ GeV$^2$,
these relations stay the same, although the distributions
themselves shift to smaller $z$ because of the $Q^2$ evolution
and the second moments slightly change.

This work is the first clear evidence on the internal configuration
of the $f_0$(980) from a global analysis of fragmentation function 
data. In addition, the $s\bar s$ configuration was indicated
in high-energy heavy-ion reactions \cite{CMS-f0-2025}.
However, it is in contradiction to many other studies
especially from low-energy studies that the $f_0$(980) should be
a tetraquark or $K\bar K$ molecule.
For example, this exotic composition was confirmed by two-photon decay 
and radiative-decay widths of $f_0$(980) \cite{pdg-2024}.
This paradox could be solved by the following consideration.
In the past, we investigated another exotic hadron candidate
$\Lambda$(1405), which is considered as a pentaquark or 
$\bar K N$ molecule, in high-energy exclusive reactions 
by using the constituent counting rule of perturbative QCD.
The $\Lambda$(1405) photo-production data suggested that 
$\Lambda$(1405) becomes an ordinary $qqq$ state at high energies
although it could be a pentaquark state at low energies.
In the same way, we can imagine that $f_0$(980) becomes 
a $q\bar q$ state at high energies although it could be
a  tetraquark state at low energies \cite{const-conting-2016}.
The current work on the FF analysis supports this picture:
\vspace{-0.20cm}
\begin{align}
f_0 (980) =  \ \text{tetraquark state (at low energies)}
  \ \to \  q\bar q \ \text{state (at high energies)}
\nonumber
\label{eqn:f0-low-high-energies}
\vspace{-0.00cm}
\end{align}
\\[-1.20cm]

Namely, the internal configuration changes by the energy we look
at the hadron.
If this explanation is valid, a new hadron physics field
could be created. Exotic hadron candidates could be considered
as ``exotic" hadrons at low energies; however, they
could be viewed as ordinary hadrons 
with the $q\bar q$ and $qqq$ configurations at high energies.
In order to establish this transition picture, 
experimental confirmations are needed
by changing reaction energies or momenta.
The original naive quark model could be valid at high energies,
where quarks and gluons are explicit fundamental degrees 
of freedom of QCD, whereas some hadrons are considered as
exotic ones at low energies.

\vspace{-0.10cm}
\section{Summary}
\label{summary}
\vspace{-0.10cm}

A global analysis has been done for the FFs of an exotic hadron 
candidate $f_0$(980). From the determined second moments of the FFs
and their $z$-dependent functional forms, we concluded that 
$f_0$(980) should have the $s\bar s$ configuration, which is 
in contradiction to many studies at low energies to indicate 
a tetraquark or $K\bar K$ configuration.
This issue could be resolved by considering that
the internal hadron structure changes depending on the examined energy.
Namely, exotic hadrons could be viewed as exotic at low energies,
but they could become ordinary $q\bar q$ and $qqq$ hadrons at high energies,
where quarks and gluons are explicit degrees of freedom. 
This kind of new idea should be tested by future experiments
especially by looking at reaction energy or momentum dependencies.

\vspace{-0.10cm}



\begin{thebibliography}{99}
\setlength{\itemsep}{-0.01cm}
\setlength\baselineskip{14pt}
\vspace{-0.22cm}
\bibitem{pdg-2024}
   For a recent review on $f_0 (980)$, see
   S. Navas {\it et al}. (Particle Data Group), 
   \href{https://doi.org/10.1103/PhysRevD.110.030001}
      {\emph{Phys. Rev. D} {\bf 110} (2024) 030001},
   especially Sec.\,64, 
   Scalar Mesons below 1 GeV, for finding the current
   situation of possible $f_0 (980)$ configurations.
\bibitem{cik-1993}
   F. E. Close, N. Isgur, S. Kumano,
   \href{https://doi.org/10.1016/0550-3213(93)90329-N}
      {\emph{Nucl. Phys. B} {\bf 389} (1993) 513}.
\bibitem{sk-2015}      
   T. Sekihara and S. Kumano,
   \href{https://doi.org/10.1103/PhysRevD.92.034010}
      {\emph{Phys. Rev. D} {\bf 92} (2015) 034010}.     
\bibitem{gpd}
   H. Kawamura and S. Kumano,
   \href{https://doi.org/10.1103/PhysRevD.89.054007}
      {\emph{Phys. Rev. D} {\bf 89} (2014) 054007}.
\bibitem{const-conting-2013}
   H. Kawamura, S. Kumano, and T. Sekihara, 
   \href{https://doi.org/10.1103/PhysRevD.88.034010}
      {\emph{Phys. Rev. D} {\bf 88} (2013) 034010}.
\bibitem{const-conting-2016}
   W.-C. Chang, S. Kumano, and T. Sekihara, 
   \href{https://doi.org/10.1103/PhysRevD.93.034006}
      {\emph{Phys. Rev. D} {\bf 93} (2016) 034006}.
\bibitem{hkos-2008} 
   M. Hirai, S. Kumano, M. Oka, and K. Sudoh,
   \href{https://doi.org/10.1103/PhysRevD.77.017504}
      {\emph{Phys. Rev. D} {\bf 77} (2008) 017504}.
\bibitem{sk-ffs-2025}
   S. Kumano, 
   \href{https://doi.org/10.48550/arXiv.2511.20217}
       {arXiv:2511.20217}, Phys. Rev. D (Letter) in press.
\bibitem{f0width-th}
    S. Kumano and V. R. Pandharipande, 
    \href{https://doi.org/10.1103/PhysRevD.38.146}
      {\emph{Phys. Rev. D} {\bf 38} (1988) 146}.  
\bibitem{KEKB-2025}
   R. Seidl {\it et al.} (Belle Collaboration), 
   \href{https://doi.org/10.1103/PhysRevD.111.052003}
      {\emph{Phys. Rev. D} {\bf 111} (2025) 052003}.
\bibitem{esw-book} 
    R. K. Ellis, W. J. Stirling, and B. R. Webber,
      {\it QCD and Collider Physics}, Cambridge University Press (1996).
\bibitem{FFs-global-1}
   M. Hirai, S. Kumano, T.-H. Nagai, and K. Sudoh,
   \href{https://doi.org/10.1103/PhysRevD.75.094009}
      {\emph{Phys. Rev. D} 75 (2007) 094009};
   M. Hirai, H. Kawamura, S. Kumano, and K. Saito,  
   \href{https://doi.org/10.1093/ptep/ptw154}
      {\emph{Prog. Theor. Exp. Phys.} {\bf 2016} (2016) 113B04}.
\bibitem{FFs-global-2}
   There are many global analyses for light hadrons.
   For a complete list of FFs analyses, see the recent paper
   J. Gao {\it et al.},
   \href{https://doi.org/10.1103/t5ds-vvc4}
      {\emph{Phys. Rev. D} {\bf 112} (2025) 054045}.
\bibitem{FFs-evol-2012}
   M. Hirai and S. Kumano, 
   \href{https://doi.org/10.1016/j.cpc.2011.12.022}
      {\emph{Comput. Phys. Commun.} {\bf 183} (2012) 1002}.
\bibitem{hrs86}
   S. Abachi {\it et al.} (HRS Collaboration), 
   \href{https://doi.org/10.1103/PhysRevLett.57.1990}
      {\emph{Phys. Rev. Lett.} {\bf 57} (1986) 1990}.    
\bibitem{delphi9599}
   P. Abreu {\it et al.} (DELPHI Collaboration), 
   \href{https://doi.org/10.1007/BF01578668}
      {\emph{Z. Phys. C} {\bf 65} (1995) 587};
   \href{https://doi.org/10.1016/S0370-2693(99)00105-7}
      {\emph{Phys. Lett. B} {\bf 449} (1999) 364}.
\bibitem{opal98}
   K. Ackerstaff {\it et al.} (OPAL Collaboration), 
   \href{https://doi.org/10.1007/s100529800886}
      {\emph{Eur. Phys. J. C} {\bf 4} (1998) 19}.   
\bibitem{CMS-f0-2025}
   A. Hayrapetyan {\it et al.} (CMS Collaboration), 
   \href{https://doi.org/10.1038/s41467-025-56200-6}
      {\emph{Nature Commu.} {\bf 16} (2025) 7990}.
\end{thebibliography}
\end{document}